\documentclass{article} 
\usepackage{bris}
\usepackage{psfig}  \usepackage{epsfig}
\def\lsim{\mathrel{\rlap{\lower4pt\hbox{\hskip1pt$\sim$}}
    \raise1pt\hbox{$<$}}}                
\def\gsim{\mathrel{\rlap{\lower4pt\hbox{\hskip1pt$\sim$}}
    \raise1pt\hbox{$>$}}}                
\newcommand{\ee}{\mbox{$e^+e^-$}}
\newcommand{\la}{\langle}
\newcommand{\ra}{\rangle}
\begin{document}
\begin{titlepage}{BRIS/HEP/2000-08}{GLAS-PPE/2000-11}{November
2000}
\title{\bf A Comparison of the Momentum
Spectra in Deep Inelastic Scattering with the Modified Leading Logarithmic
Approximation.}
\author{N.~H.~Brook\Instref{bris} I.~O.~Skillicorn\Instref{glas}}
\Instfoot{bris}{H. H. Wills Physics Lab., University of Bristol, UK.}
\Instfoot{glas}{Dept. of Physics \& Astronomy, University of Glasgow, UK.}
\begin{abstract}
The evolution of the logarithmic scaled energy
spectra with energy scale, in the framework of the modified
leading logarithmic approximation (MLLA), is investigated.
The behaviour of the higher order moments
of HERA deep inelastic scattering data
are compared with the analytic predictions of MLLA.
\end{abstract}
\end{titlepage}

\section{Introduction}
The properties of the hadronic final state cannot be
calculated directly in Quantum Chromodynamics (QCD).
It is necessary to
model the transition from the perturbatively calculable partonic final state
to hadrons.  One approach that has been successful
in describing the general features of the inclusive energy spectra,
in both $\rm e^+e^-$ annihilation and deep inelastic
scattering (DIS) experiments~\cite{JETP,MLLArev},  is the modified leading log
approximation (MLLA). This uses the local parton hadron duality (LPHD) 
approach~\cite{LPHD}
to relate the MLLA partonic predictions to the hadronic
observables.

The hadronic final state reflects the underlying partonic behaviour. 
At small values of momentum fraction carried by the outgoing
parton from the original hard
scatter, the parton  
shower development is dominated by gluon bremsstrahlung. 
The  
partonic properties can be calculated, on average, in the framework of
the MLLA. These MLLA calculations have
two free parameters: a running strong coupling, defined by
a QCD scale $\Lambda,$ and an energy cut-off, $Q_0,$ below which the
parton evolution is truncated.

LPHD assumes that a simple normalisation factor,
which does not explicitly include either the effect of particle masses
or resonance decays,
 can be used to
relate the hadronic distributions to the partonic ones.
At large enough energies, away from where the parton evolution is
truncated, this `hadronisation' factor should be approximately
independent of the energy scale~\cite{LPHD}
 at which the parton spectra are calculated.

In our previous study~\cite{prev_paper},
various approaches were taken to calculate
the single particle spectra and their moments within the MLLA framework
and compared with $\rm e^+e^-$ data.
In this paper the theoretical results
are compared with deep inelastic scattering data
in the current region of the Breit frame~\cite{ZEUS}.

The event kinematics of DIS are determined by the negative square of the 
four-momentum of the virtual exchanged boson,
$Q^2\equiv-q^2$, and the Bjorken scaling variable, $x=Q^2/2P\!\cdot\!q$,
where $P$ is the four-momentum of the proton.
In the Quark Parton Model (QPM),
the interacting quark from the proton carries four-momentum $xP.$

A natural frame in which to study the dynamics of the hadronic final
state
in DIS is the Breit frame~\cite{feyn}.
In this frame, the exchanged
virtual boson is completely space-like and has a four-momentum
$q = (0,0,0,-Q=-2xP^{Breit})$,
where $P^{Breit}$ is the momentum of the proton in the Breit frame.
The particles produced in the 
interaction can be assigned to one of two regions, known as the
current and target regions.
The Breit frame allows the
maximal  separation of the incoming and outgoing partons
in the QPM. 

The current region in the Breit frame of DIS
is analogous to a single hemisphere of $e^+e^-$ annihilation.
In the current region, 
the outgoing struck quark from the proton
has a momentum $-Q/2.$ 
The scaled momentum spectra of the particles in the current region, 
expressed in terms of $\xi_p = \ln(Q/2p^{Breit}),$
are expected to have a
dependence on $Q$ similar to that observed in
\ee~annihilation~\cite{JETP} at energy $\sqrt{s_{ee}}=Q.$
It should be noted though, that the current region of DIS will have a
different primary quark flavour composition than $\rm e^+e^-$
annihilation, in particular for heavy quarks.

This paper extends the original ZEUS analysis by investigating the
effect of introducing an effective mass term into the MLLA predictions
(see section~\ref{sec:theory}.)
In contrast to the previous analysis of $\rm e^+e^-$ data~\cite{prev_paper} 
this effective mass was treated as an additional free parameter of the
theoretical models and was fitted to the data.

\section{Single Particle Spectra}
\label{sec:theory}
Given a high-energy parton that fragments via secondary partons
into a jet of hadrons,
the MLLA evolution equation allows the secondary parton spectra for the
logarithmic scaled energy, $\xi,$ to be calculated~\cite{dokevol}. 
The variable $\xi$ is
defined as $\ln(E_0/E) \equiv \ln(1/x),$ 
where $E_0(=Q/2$ for DIS)
is the energy of the original parton 
and $E$ is the energy of the secondary 
parton. The cut-off, $Q_0,$ limits the parton
energy to $ E \ge k_T \ge Q_0,$ where $k_T$ is the transverse energy of
the decay products in the jet evolution. In order
to reconstruct the $\xi$ distributions
an inverse Mellin transformation is performed.

It is convenient to investigate the MLLA spectra in terms of moments.
The analytic form of the cumulant moments has been calculated 
in~\cite{dokevol} for the first four moments. This allows the
normalised moment, $\langle \xi^m \rangle,$ to be calculated and hence
the dispersion ($\sigma$), skewness ($s$) and the kurtosis ($k$) 
to be constructed, see for example ~\cite{kendall} 

\begin{eqnarray}
\langle \xi \rangle & = & K_1  \\
\nonumber \\
\sigma^2 & = & K_2  = \langle\xi^2\rangle - \langle\xi\rangle^2 \\ 
\nonumber \\
s & = & K_3/\sigma^3  = 
\frac{\langle\xi^3\rangle-3\langle\xi^2\rangle\langle\xi\rangle+2\langle\xi
\rangle^3}{\sigma^3} \\ 
\nonumber \\
k & = & K_4/\sigma^4  = 
\frac{\langle\xi^4\rangle-4\langle\xi^3\rangle\langle\xi\rangle-
3\langle\xi^2\rangle^2+12\langle\xi^2\rangle\langle\xi\rangle^2
-6\langle\xi\rangle^4}
{\sigma^4}.
\end{eqnarray}

The so-called limiting spectrum~\cite{LPHD}
 is the case when $Q_0=\Lambda.$
The $\xi$ distribution, $\bar D(\xi,Y),$ is dependent on one
parameter $Y,$ which is given by $\ln(E_0/\Lambda),$ where $\Lambda$ is
now the only free parameter.
For this limiting case, Fong and Webber~\cite{fongweb} have
also calculated the behaviour of the moments of the $\xi$
spectra with energy scale $Q.$
They point out that the spectra
 can be represented close to the
maximum of the distribution  
by a distorted Gaussian of the form:

\begin{equation}
\bar D(\xi,Y) \propto \exp\left[ \frac{1}{8}k-\frac{1}{2}s\delta
-\frac{1}{4}(2+k)\delta^2 + \frac{1}{6}s\delta^3
+\frac{1}{24}k\delta^4 \right]
\label{eq:distg}
\end{equation}
where  $\delta = (\xi - l)/\sigma$ and $l$
is the mean. 
The differences between   the parton spectra for quark (and gluon) jets
are Next-to-MLLA effects and have been included in these investigations
according to the approximate prescription of Dokshitzer et al.~\cite{pQCD}.




The MLLA  limiting spectra predicts that the $E_0$-dependence of 
$\xi_{\rm max}$ should have the form 

\begin{equation}
\xi_{\rm max} = Y\left(0.5 + \sqrt\frac{C}{Y} -\frac{C}{Y} +
\left(\frac{d}{Y}\right)^{-\frac{3}{2}} \right),
\label{eq:maxevol}
\end{equation}

\noindent where 
$C$ is expected to be $\approx 0.3$~\cite{pkevol}.

The mean and peak position, $\xi_{\rm max},$ are related,
for the skew Gaussian used here, by

\begin{equation}
\xi_{\rm max} = \la \xi \ra - \frac{s}{\sigma}+\frac{\sigma}{s}
(1+s^2)^{\frac{1}{2}}.
\label{eq:maxmean}
\end{equation}

\noindent Thus the maximum is a function of several variables.

In the MLLA approach discussed above, 
the partons are assumed massless so the scaled energy 
and momentum spectra are identical. Experimentally the scaled momentum
distribution is usually measured and as the observed hadrons are massive the
equivalence of the two spectra no longer holds. In~\cite{klo} the assumption
is made that the cut-off $Q_0$ can be related to an effective mass,
$m_{\rm eff},$ of the hadrons.
This allows the logarithmic
scaled momentum distribution, $\xi_p,$ to be written as

\begin{equation}
\frac{1}{N}\frac{dn_h}{d\xi_p} \propto
\frac{p_h}{E_h}\bar D(\xi,Y), 
\label{eq:xip}
\end{equation}
where 
$$
\xi = \log\frac{E_0}{\sqrt{E_0^2e^{-2\xi_p}+m^2_{\rm eff}}}
$$
\noindent and the energy of a hadron with a momentum $p_h$ is 
$E_h = \sqrt{p_h^2+m_{\rm eff}^2}.$
The limiting momentum spectra based on massless partons or massive partons 
with the $m_{\rm eff}$ being set to $Q_0 \equiv \Lambda,$ will
be referred to as MLLA-0 and MLLA-M spectra, respectively. 

\section{Comparison with DIS Data}
\begin{figure}[t]
\centerline{\psfig{figure=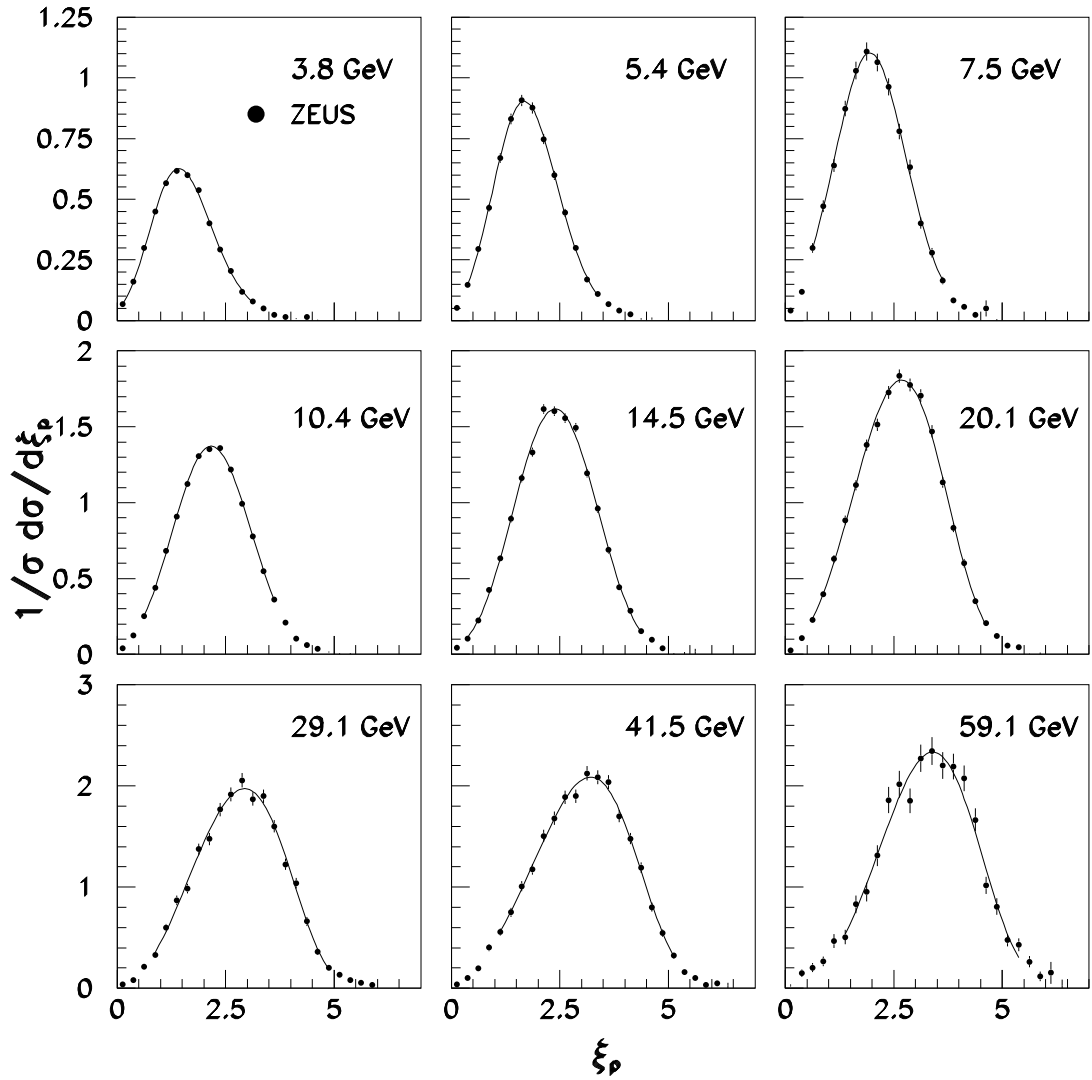,width=0.95\textwidth}}
\caption{The charged particle distributions of
$1/N dn /d\xi_p$ in the current fragmentation
region as a function of $\xi_p$
for different $Q$ bins.
Only statistical errors are shown. The full line is the skew
Gaussian fit.}
\label{fig:logxp}
\end{figure}

Figure~\ref{fig:logxp} shows the differential
$\xi_p$ distributions, with statistical errors only,
for charged particles in the current fragmentation
region of the Breit frame for nine 
$\la Q \ra$ values from the ZEUS experiment~\cite{ZEUS}.
These distributions are approximately Gaussian in shape with
the mean charged multiplicity given by the integral of the distributions.
As $Q$ increases, the multiplicity increases and, in addition,
the peak of the
distribution moves to larger values of $\xi_p.$
The data were fitted to the
distorted Gaussian of eqn.~\ref{eq:distg}
over a range of $\pm1.5$ units (for $\la Q \ra< 14.5 {\rm\ GeV}$) or 
$\pm2$ units (for $\la Q \ra \ge 14.5 {\rm\ GeV}$) in $\xi_p$ around the mean.
This is the range used by the ZEUS collaboration in their analysis
of the data.
The smooth curves in Fig.~\ref{fig:logxp} result from the fit
to the data;
they represent the data well.
The fitted parameters were insensitive
to the range of the fit.
The $\chi^2$ values are not quoted here because the measurements are
dominated by systematic errors.

\begin{figure}[t]
\centerline{\psfig{figure=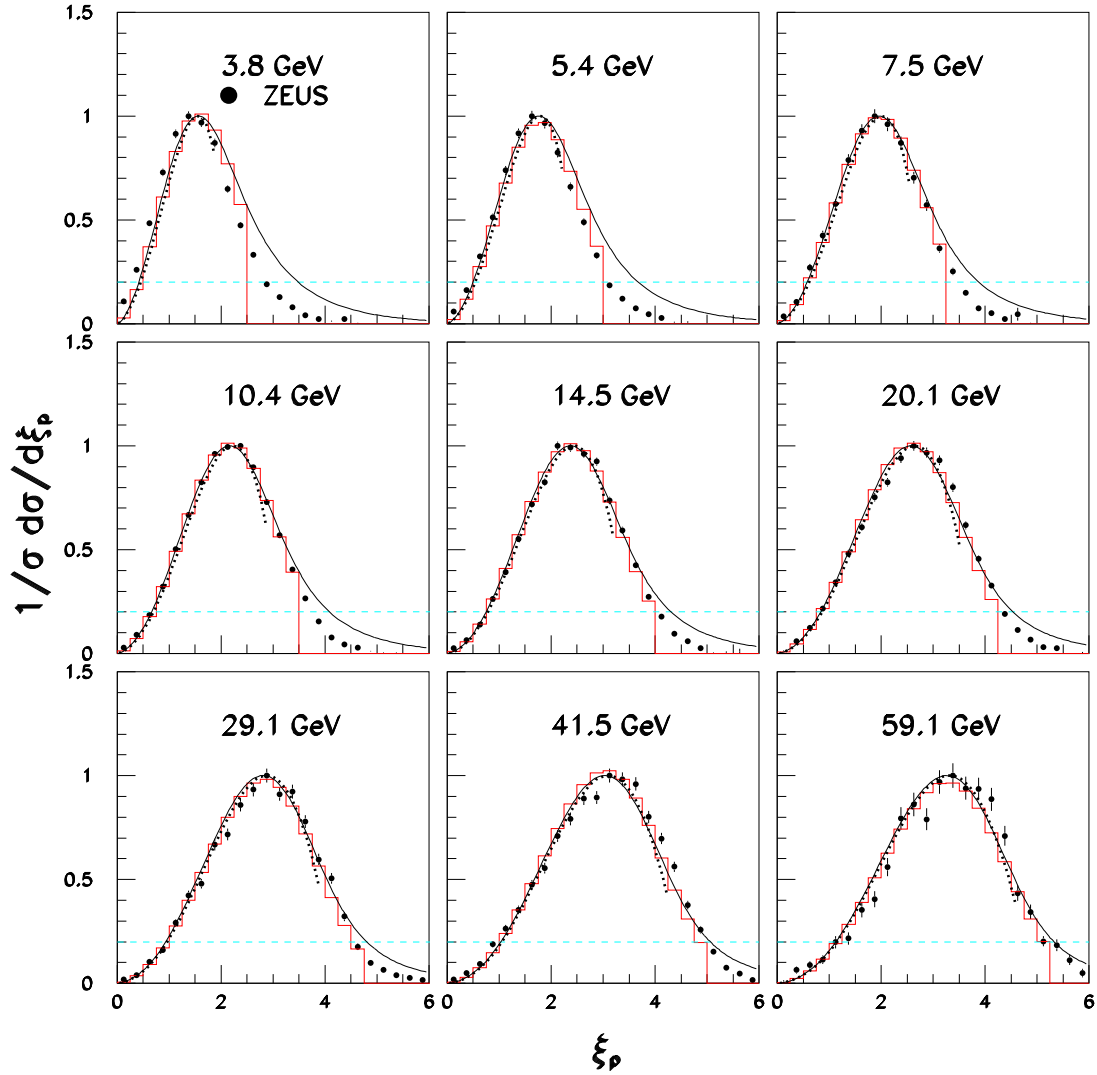,width=0.95\textwidth}}
\caption{The charged particle distributions of
$1/N dn /d\xi_p$ in the current fragmentation
region as a function of $\xi_p$
for different $Q$ bins.
Only statistical errors are shown. The full and dotted lines are the
MLLA-M and MLLA-0 predictions with $\Lambda=280 {\rm\ MeV}$
respectively. The histogram is fit of the MLLA to the data with
$\Lambda = 280\pm 3 {\rm\ MeV}$ and $m_{\rm eff} = 230\pm 8 {\rm\ MeV}.$ }
\label{fig:datamlla}
\end{figure}

The ZEUS data, solid points, are compared
with various MLLA curves in Figure~\ref{fig:datamlla}. 
For clarity both the data and theory are normalised
to a maximum value of unity. The theoretical curves (solid line 
MLLA-M, dotted line MLLA-0) are calculated for $\Lambda = 280 {\rm\ MeV}
(=m_{\rm eff}).$
Both the MLLA-0 and MLLA-M describe well the 
general trend of the data. In particular, MLLA-M describes well the
central region of data for $\la Q\ra  > 10.4 {\rm\ GeV.}$ There is a trend
however, for the MLLA-M distribution to be more positively skew 
than the data; this is particularly pronounced at low energies.
In contrast, MLLA-0 is consistently less skew than the data
with the skewness being negative at low $Q.$ MLLA-M and MLLA-0
converge to similar distributions as $Q$ increases.
At low $Q$ the data are peaked at lower values of $\xi_p$ than the
theoretical predictions while at high $Q$ the predictions peak
lower than the data.

The histogram is a fit of MLLA, in the theoretically allowed region,
 over the range of data above the
horizontal dashed line (80\% of the data 
as measured from the maximum of the distributions), 
taking both $\Lambda$ {\bf and}
$m_{\rm eff}$ to be free parameters.
The normalisation of the distribution in each $Q$ bin was allowed to
vary.
The fit gives $\Lambda = 280\pm 2 {\rm\ MeV}$ and
$m_{\rm eff} = 230\pm 6 {\rm\ MeV}.$
Again, within the errors, the parameters were insensitive
to the region of the fit to the data.
The fitted MLLA gives a distribution intermediate between 
MLLA-M and MLLA-0. The data are well fitted for $\la Q\ra  > 10.4 {\rm\ GeV.}$
Otherwise, the fit closely follows MLLA-M and shows a similar 
disagreement with the data.

\begin{figure}[t]
\centerline{\psfig{figure=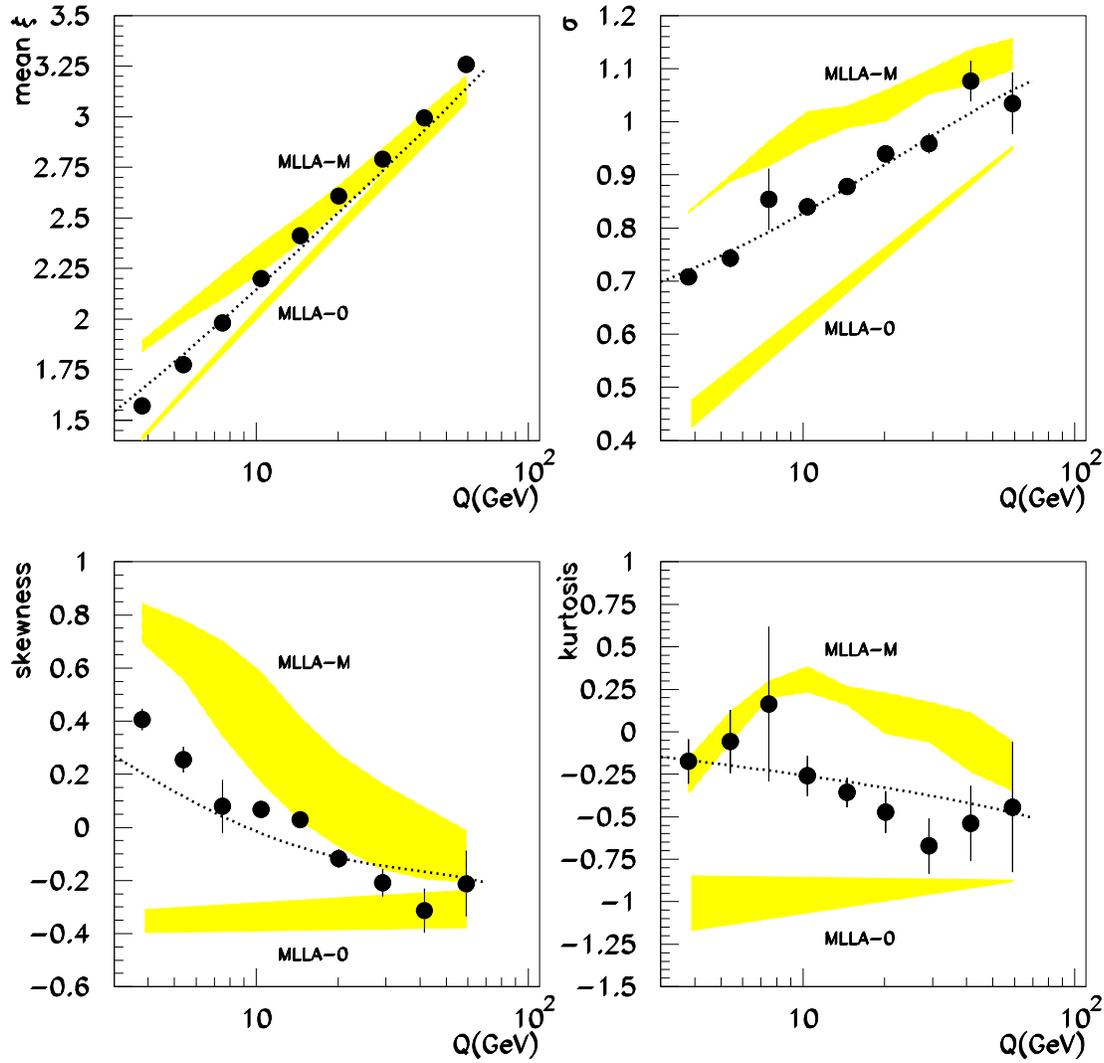,width=0.95\textwidth}}
\caption{The moments of the 
$\xi_p$ distribution in the current fragmentation
region as a function of $Q$, from fits to ZEUS data, are 
shown as solid points.
The errors  are those returned from the fit. 
The upper and
lower shaded areas are from fits to the MLLA-M and MLLA-0
spectra, respectively. 
The dotted line shows the moments obtained from a fit   
to the MLLA spectrum obtained using the fitted $\Lambda  = 280 {\rm\ MeV},
m_{\rm eff} = 230 {\rm\ MeV.}$}
\label{fig:evol}
\end{figure}

To study further the differences between the MLLA predictions
and the data, fits of a skew Gaussian have been made to
the MLLA predictions, in a manner similar to the fits made
to the data, in order to extract the moments.
Figure~\ref{fig:evol} shows the first 
four moments of the $\xi_p$ spectra as a function of $Q.$
For the ZEUS data (solid points), the mean and $\sigma$ increase as
$Q$ increases, whilst the skewness and kurtosis fall.
Also shown are the results from the fits to
the MLLA-M and MLLA-0 spectra, 
the upper and lower shaded areas respectively. For both MLLA-0 and
MLLA-M the range fitted
lay between 30\% of the spectrum (lower limit of band) to
80\% of the spectrum (upper limit of band). 
The upper limit of the fit in $\xi_p$ was determined
by the limit of the MLLA spectrum if this were less than the percentage
range.
In all cases the skew Gaussian describes the theory to better than 0.5\%.

There are two points to be noted when making
these fits: firstly the fitted parameters of the theory depend on 
the range of spectrum fitted. Secondly, fitting over a range
greater than $\pm 1\sigma$ may mean that the skewness and kurtosis
are not those of the distribution because the approximations
made in deriving the expression become invalid.  Consequently
the  the fitted parameters should be considered as simply
a parameterization of the distributions rather than accurate
measures of the skewness and kurtosis. Equally because
the $\sigma$ and kurtosis are correlated, the measured $\sigma$
is fit-range dependent. 

For $\sigma,$ skewness and kurtosis, Fig.~\ref{fig:evol},
 the DIS data are bracketed by the MLLA-M and MLLA-0 predictions
as would be expected from the plots of Figure~\ref{fig:datamlla}.
The skewness predicted by MLLA-M is more positive than the data
while MLLA-0 underestimates the skewness; for the highest $Q$
both models approach the data.
There is evidence of a systematic difference in the $Q$-dependence
of the mean between the 
models and the data for  the energy range studied here. 

The dotted line shows the moments obtained from a fit   
to the massive MLLA spectrum obtained  using the fitted $\Lambda  = 280
{\rm\ MeV}$ and $m_{\rm eff} = 230 {\rm\ MeV.}$  
It is evident that a reasonable description
of all moments other than the mean is obtained; the mean has 
an energy dependence that differs from that of the data.

\begin{figure}[t]
\centerline{\psfig{figure=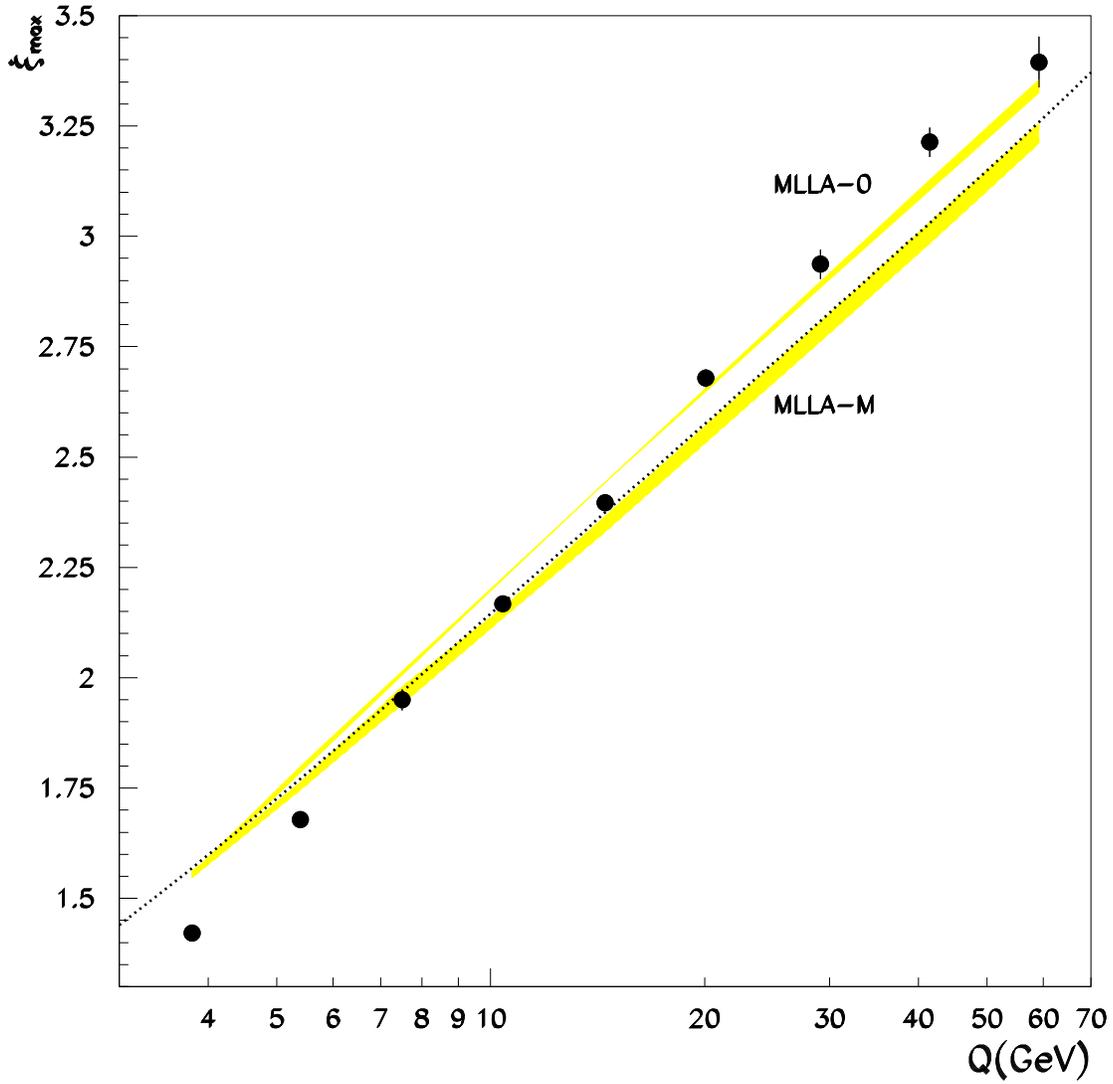,width=0.95\textwidth}}
\caption{The peak position of the 
$\xi_p$ distribution in the current fragmentation
region as a function of $Q.$
The errors shown are those returned from the fit. 
The upper and
lower shaded areas are from fits to the MLLA-M and MLLA-0
spectra, respectively. 
The dotted line shows the $\xi_{\rm max}$ obtained from a fit   
to the MLLA spectrum obtained using the fitted $\Lambda  = 280 {\rm\ MeV},
m_{\rm eff} = 230 {\rm\ MeV.}$}
\label{fig:peak}
\end{figure}

The energy dependence  of the $\xi_p$ distribution
has been explored further by studying the dependence of the
peak position on $Q.$ 
This has the advantage, in comparison with the moments, that the maximum is
relatively insensitive to the function  used to parameterize the shape
of the $\xi_p$ distribution.
The $Q$-dependence of $\xi_{\rm max}$ is shown in  Figure~\ref{fig:peak}. 
The shaded bands show the maxima derived from MLLA-0 and MLLA-M. 
For both cases the
range of fit is from 30\% to 80\%. 
As before the dotted line uses the MLLA parameters
$\Lambda = 280 {\rm\ MeV}$ and $m_{\rm eff} = 230 {\rm\ MeV}$ from the fit. 
This plot confirms that the energy dependence of the DIS
data differs from that of any of the MLLA predictions.

A good fit of eqn.~\ref{eq:maxevol}
 to the $\xi_{\rm max}$ from the experimental data can be achieved with 
$\Lambda=285 {\rm\ MeV}$ 
(with $d$ set to zero), which is 
consistent with  the $\Lambda$ value from the fit to the spectrum.
However the value of $C$ is a factor two larger than expected
and  indicates again  an inconsistency
of the energy dependence of the data with the expectations of the MLLA.

These results are consistent with our previous analysis of $\rm e^+e^-$
data~\cite{prev_paper}.

\section{Conclusions}
The HERA DIS data in the current fragmentation region of the
Breit frame have been compared 
to the MLLA predictions as a function of $Q.$ In general, 
the data are well described within the LPHD-MLLA approach, with the
predictions of MLLA-0 and MLLA-M bracketing the data. With increasing
$Q,$ the various predictions of the MLLA converge on the data.
A fit has been made 
to the data using the massive MLLA approach and values of $\Lambda$ and
$m_{\rm eff}$ are found that give a reasonable description of the 
moments of the $\xi$ distribution.

In detail discrepancies are found
between the theoretical predictions and the data.
In particular, a consistent description of the evolution of the 
peak position of the $\xi$ distribution as a function of scale $Q$ is
not possible; no energy-independent 
combination of $\Lambda$ and $m_{\rm eff}$ is found
that can describe the data.

\section*{Acknowledgements}
The authors would like to thank Valery Khoze and Wolfgang Ochs
for discussion and their many useful comments.

\newpage

\newpage

\newpage

\newpage

\end{document}